\documentclass[twocolumn,prl]{revtex4}
\usepackage{graphics,graphicx}
\begin{document}
\title{Spin waves in the Frustrated Kagom\'e Lattice Antiferromagnet KFe$_{3}$(OH)$_{6}$(SO$_{4}$)$_{2}$}
\author{K.~Matan$^{1}$}
\author{D.~Grohol$^{2}$}
\author{D.~G.~Nocera$^{2}$}
\author{T.~Yildirim$^{3}$}
\author{A.~B.~Harris$^{4}$}
\author{S.~H.~Lee$^{3,5}$}
\author{S.~E.~Nagler$^{6}$}
\author{Y.~S.~Lee$^{1}$}
 \affiliation{$^{1}$Department of Physics, Massachusetts Institute of Technology, Cambridge, MA 02139}
 \affiliation{$^{2}$Department of Chemistry, Massachusetts Institute of Technology, Cambridge, MA 02139}
 \affiliation{$^{3}$NIST Center for Neutron Research, Gaithersburg, MD 20899}
 \affiliation{$^{4}$Department of Physics and Astronomy, University of Pennsylvania, Philadelphia, PA 19104}
 \affiliation{$^{5}$Department of Physics, University of Virginia, Charlottesville, VA 22904}
 \affiliation{$^{6}$Center for Neutron Scattering, Oak Ridge National Laboratory, Oak Ridge, TN 37831}
\date{\today}

\begin{abstract}

The spin wave excitations of the ideal $S=5/2$ Kagom\'e lattice
antiferromagnet KFe$_{3}$(OH)$_{6}$(SO$_{4}$)$_{2}$ have been
measured using high-resolution inelastic neutron scattering.  We
directly observe a flat mode which corresponds to a lifted ``zero
energy mode,'' verifying a fundamental prediction for the Kagom\'e
lattice. A simple Heisenberg spin Hamiltonian provides an excellent
fit to our spin wave data.  The antisymmetric Dzyaloshinskii-Moriya
interaction is the primary source of anisotropy and explains the low
temperature magnetization and spin structure.

\end{abstract}

\pacs{75.25.+z,75.30.Ds,75.50.Ee} \maketitle

Geometrically frustrated spin systems have received considerable
attention in recent years due to the presence of remarkable
properties such as spin ice\cite{ramirez:1,schiffer:2}, spin
nematic\cite{chalker_92}, and spin liquid behaviors
\cite{waldtmann,sachdev,elser}.  The Kagom\'e lattice
antiferromagnet is a highly frustrated two-dimensional lattice,
being comprised of corner-sharing triangles.  For classical
Heisenberg spins, the ground state of a Kagom\'e antiferromagnet is
infinitely degenerate, but the spins are believed to order in the $T
\rightarrow 0$ limit by a process known as ``order by disorder''
\cite{huse:7536,reimers:9539}.  On the other hand, there are
predictions that the ground state of the $S=1/2$ Kagom\'e lattice is
disordered, being a realization of the long sought after quantum
spin-liquid\cite{waldtmann,sachdev,elser,sindzingre}.
Experimentally, several materials have been studied that are
believed to be realizations of the Kagom\'e lattice antiferromagnet,
such as SCGO\cite{broholm_SCGO}, volborthite\cite{hiroi}, and
jarosites\cite{wills,inami,shl_cr}. However, these materials are
often plagued by non-stoichiometry issues or have structural
differences from the ideal Kagom\'e network.  In this paper, we
present a high-resolution neutron scattering study on a single
crystal sample of a pure system with an ideal Kagom\'e
network\cite{grohol}, the iron jarosite
KFe$_{3}$(OH)$_{6}$(SO$_{4}$)$_{2}$. This allows us to directly
compare our data with fundamental theoretical predictions.

One of the hallmarks of highly frustrated systems is the presence of
``zero energy modes'' which result from the huge ground-state
degeneracy.  For the Kagom\'e lattice Heisenberg model, the only
constraint for the ground state is that the spins on each triangle
be oriented $120^\circ$ relative to each other. A ``zero energy
mode'' for the Kagom\'e lattice is depicted in Fig.~1(a). The small
loops at the tips of the spins illustrate rotations of two of the
spin sublattices about the axis defined by the third spin
sublattice.  These spins, forming a chain, can collectively rotate
around the loop paths with no change in energy (the 120$^\circ$
angles on each triangle are maintained). Furthermore, the spins on
different parallel chains can be excited independently.  Hence, this
type of excitation costs no energy and is
non-dispersive\cite{harris}.  This mode has not been directly
observed previously, and, since it occurs at zero energy, it is
difficult to measure with most experimental techniques. Here, we
report a direct observation of a zero energy mode in iron jarosite,
which is lifted to a finite energy due to the presence of spin
anisotropy resulting from the Dzyaloshinskii-Moriya (DM)
interaction\cite{dzyaloshinskii,moriya}.

KFe$_{3}$(OH)$_{6}$(SO$_{4}$)$_{2}$ is an ideal realization of a
Kagom\'e Heisenberg antiferromagnet due to its undistorted lattice,
fully occupied magnetic sites, and weak interlayer
coupling\cite{grohol,grohol_chiral}.  The $S=5/2$ Fe$^{3+}$ ions
form a perfect Kagom\'e arrangement and are surrounded by an
octahedral environment of oxygens.  Previous powder neutron
diffraction studies\cite{inami,wills} identified the long-range
magnetic order below $T_N=65$ K as that shown in Fig.~1(a). For this
spin arrangement, each triangle has positive vector
chirality\cite{inami,grohol_chiral}, such that the spins rotate
clockwise as one traverses the vertices of a triangle clockwise.
Magnetization results on single crystals have determined that the
spins are canted slightly out of the Kagom\'e plane, yielding an
``umbrella'' structure. Moreover, each plane has a net ferromagnetic
moment, with an antiferromagnetic coupling between adjacent
planes\cite{grohol_chiral}.

\begin{figure}
\centering \vspace{-2mm}
\includegraphics[width=8.0cm]{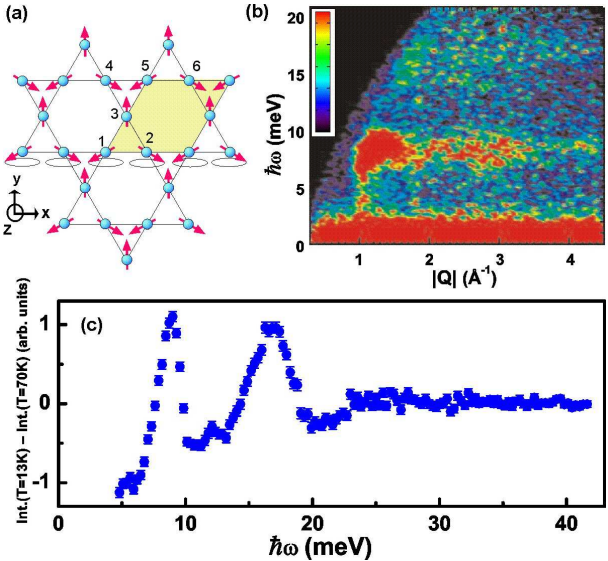}
\caption{(color online) (a) The ground state spin configuration of
iron jarosite.  The magnetic unit cell, shown by the yellow shaded
area, is the same as the chemical unit cell. The dotted loops
illustrate the zero energy excitations as described in the text. The
Cartesian axes are used in the discussion of the DM interaction. (b)
Intensity contour map of the inelastic scattering spectrum at
$T=4$~K of a powder sample measured using the time-of-flight DCS
spectrometer with an incident neutron wavelength of 1.8 \AA. (c)
Inelastic neutron scattering measured on a powder using the BT4
spectrometer with collimations $40^\prime-20^\prime$. The data show
the difference between the intensities above and below the N\'eel
temperature $T_N=65$~K and is a measure of the spin wave density of
states. }\vspace{-4mm}
\end{figure}

We first studied the magnetic excitations using a deuterated powder
sample (${\rm mass}=4.92$~g) on the DCS and BT4 spectrometers at the
NIST Center for Neutron Research, as shown in Figures 1(b) and 1(c).
Figure~1(c) shows a difference plot of the intensity as a function
of neutron energy loss measured above ($T=70$~K) and below
($T=13$~K) the N\'eel temperature.  This difference plot removes
most of the phonon contributions to the spectrum, hence yielding the
spin wave density of states.  Despite the powder average, the
spectrum shows one sharp feature at $\hbar \omega_0 \sim 8$ meV and
a second broad peak at about $2\omega_0$. Both features appear as
excitation bands over a wide range of $|\vec{Q}|$, as shown in
Fig.1(b).  This behavior is quite similar to that observed in
strongly frustrated spinel systems where the excitation at
$\omega_0$ has been described as a local resonance\cite{shl_prl}. It
is tempting to identify the features observed in the excitation
spectrum as one- and two-magnon scattering since strong multi-magnon
scattering might be expected due to the strong frustration and cubic
terms in the spin-Hamiltonian. However, as shown below, our single
crystal measurements provide much greater detail and demonstrate
that these are regular spin wave modes, albeit with unusual
dispersive behavior.

The spin wave dispersions were obtained from inelastic neutron
scattering measurements on a single crystal sample (composed of four
co-aligned crystals of total mass 101 mg) grown using a hydrothermal
method reported previously\cite{grohol,grohol_chiral}.
High-resolution measurements were performed using the triple-axis
spectrometer HB1 at the High Flux Isotope Reactor at Oak Ridge
National Laboratory with the sample aligned in the (HK0) and (HHL)
zones with the final energy fixed at either 13.6 meV or 14.7 meV.
Vertically focused pyrolytic graphite (PG) crystals were used to
monochromate and analyze the incident and scattered neutron beams
using the $(0~0~2)$ reflection.  Horizontal collimations of
$48'-60'-\mbox{sample}-40'-120'$ were employed, and PG filters were
placed in the scattered beam to reduce higher-order contamination.
The sample was cooled to $T=10$ K using a closed cycle $^4$He
cryostat.

\begin{figure}
\centering \vspace{-2mm}
\includegraphics[width=8.5cm]{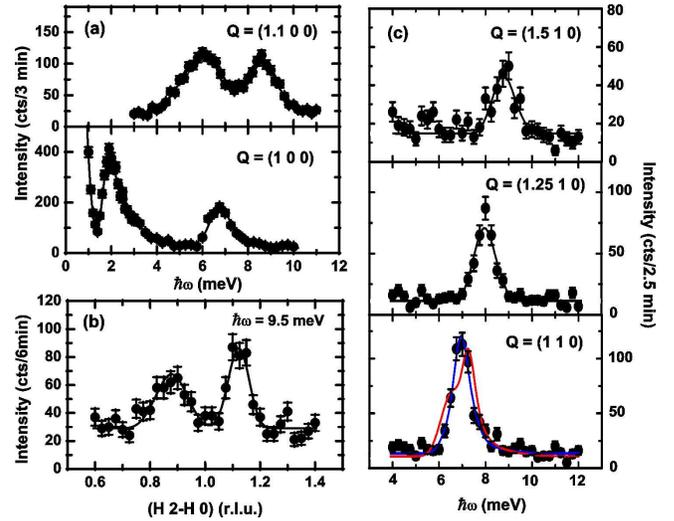}
\caption{(color online) (a) Energy scans at $\vec{Q}=(1~0~0)$ and
$(1.1~0~0)$. (b) $\vec{Q}$-scan at $\hbar \omega = 9.5$~meV. (c)
Energy scans at $\vec{Q}=(1~1~0)$, $(1.25~1~0)$, and $(1.5~1~0)$.
The solid lines show the fits to the spin wave dispersion relation
described in the text, convoluted with the instrumental resolution
function. In the lower panel of (c), the CF prediction is shown by
the red line and the DM prediction by the blue line.} \vspace{-4mm}
\end{figure}

A series of energy scans (at constant $\vec{Q}$) and $\vec{Q}$-scans
(at constant energy) were performed, and a few representative scans
are shown in Fig.~2.  The observed peaks were initially fit with
narrow Gaussians convoluted with the experimental resolution
function. Subsequent fits were performed taking into account the
empirical dispersion of the excitations.  The peaks are
resolution-limited, and the line-shapes are simply governed by the
convolution with the instrumental resolution.  A summary of all of
the peak positions and intensities is shown in Figure~3(a).  The
error bars plotted in Figs.~3(a) and (b) correspond to three times
the statistical error or one-tenth of the instrumental resolution,
whichever is larger.  The most striking feature of the data is the
relatively flat mode near 7 meV which barely disperses, even out to
the zone boundary.

The energy scan in Fig.~2(a) at the magnetic Brillouin zone center
$(1~0~0)$ reveals two spin gaps, one at 1.8(1) meV (which is
non-degenerate), and the other at 6.7(1) meV (which is two-fold
degenerate within the experimental resolution).  At
$\vec{Q}=(1.1~0~0)$, the lower-energy mode has dispersed to higher
energy and merges with the flat mode located around 7 meV.  The
other upper-energy mode disperses strongly, moving to a high zone
boundary energy of about 19 meV.  Figure~2(c) shows constant-Q scans
of the flat mode within a Brillouin zone centered at $(1~1~0)$. This
excitation barely disperses, starting from about 7 meV at the zone
center and reaching about 9 meV at the zone boundary.  We identify
this flat mode as the ``zero energy mode'' of the Kagom\'e lattice
which is lifted in energy for reasons discussed below.

\begin{figure}
\centering \vspace{0mm}
\includegraphics[width=8.7cm]{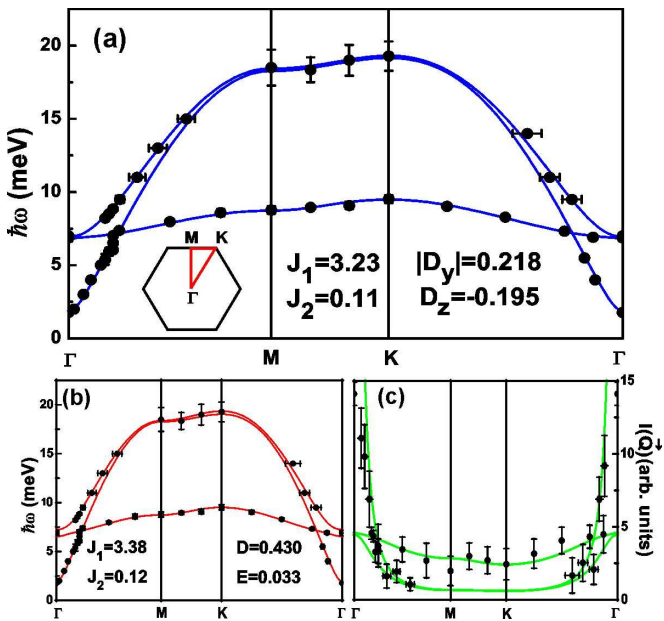}
\caption{(color online) Spin wave dispersion along the high symmetry
directions in the 2D Brillouin zone.  As discussed in the text, the
solid lines denote a fit to the DM model (in a) and to the CF model
(in b).  The model parameters are given in units of meV.  (c) Wave
vector dependence of the spin wave intensities. The solid lines
correspond to $(n(\omega)+1)/\omega(\vec{Q})$, where $n(\omega)$ is
the Bose occupation factor and $\omega(\vec{Q})$ is obtained from
the DM model. The data take into account the de-convolution with the
instrumental resolution and the Fe$^{3+}$ magnetic form
factor.}\vspace{-4mm}
\end{figure}

We have fit the observed spin wave dispersions using the following
generic Hamiltonian:\cite{yildirim}
\begin{eqnarray}
{\cal H} &=& \sum_{nn} \biggl[ J_1 \, \vec{S}_i \cdot \vec{S}_j \, +
\, \vec{D}_{ij}\cdot S_i \times S_j \biggr] \, + \, \sum_{nnn} J_2
\, \vec{S}_k \cdot \vec{S}_l
\nonumber \\
& & \, + \, \, D \sum_{i} (S_i^{y'})^2 \, - \, E \sum_{i} [
(S_i^{z'})^2 -(S_i^{x'})^2 ] \label{model_H}
\end{eqnarray}
\noindent where $\Sigma_{nn}$ ($\Sigma_{nnn}$) indicates summation
over pairs of nearest neighbors (next nearest neighbors),
$\vec{D}_{ij} = [0,D_y(i,j),D_z(i,j)]$ is the DM vector for bond
$i-j$ as shown in Fig.~1(a), and the single-ion anisotropy terms are
those used by Nishiyama {\it et al.}\cite{nishiyama} in their
treatment of the spin wave spectrum in jarosites.  Here the primed
spin components refer to the local axis associated with the rotated
oxygen octahedra (see Ref.~\onlinecite{nishiyama} for details).  We
ignore the weak interplane coupling, which is several hundred times
smaller than $J_1$\cite{grohol_chiral}.

The DM interaction is allowed for this crystal structure and merits
further discussion\cite{elhajal}. For bond 1-2, it has the form
$\vec{D}_{1,2}= (0,D_y,D_z)$. The other DM vectors can be obtained
from symmetry, such as $D_z(1,2)=D_z(2,3)=D_z(3,1)$,
$D_y(4,5)=-D_y(1,2)$ and $D_z(4,5)=-D_z(1,2)$.  Note that the
direction of the DM vector oscillates from bond to bond along the
$x$-direction.  The $z$-component of the DM vector favors the spins
to lie in the $ab-$plane and therefore effectively acts like an
easy-plane anisotropy.  The sign of $D_z$ breaks the symmetry
between positive and negative vector chirality.  Similarly the $D_y$
component breaks the rotational symmetry around the $c$-axis and
creates a small anisotropy with respect to in-plane orientations.
The effect of $D_y$ is also to cant the spins so that they have a
small out-of-plane component, consistent with the observed umbrella
spin configuration.

We may describe the spin wave data in terms of two simple spin
models. In the first of these, which we call the DM model, we
neglect the single ion anisotropy, so that the only nonzero
parameters are $J_1$, $J_2$, $D_y$, and $D_z$. In the second model,
which we call the CF (crystal field) model, all the anisotropy is
attributed to the single-ion crystal field, so that the only nonzero
parameters are $J_1$, $J_2$, $D$, and $E$. In both cases, $J_1$ is
the dominant interaction.  The numerical results obtained from these
two models are plotted as the lines in Fig.~3, and the approximate
analytic expressions for the spin gaps at the $\Gamma$ point are
given in Table~\ref{OMEGA}.

To account for the observed umbrella spin structure, we considered
the effect of spin-canting on the spin wave energies. We find that
the splitting of the mode energies at the high symmetry points are
particularly sensitive to the magnitude of the spin canting angle
out of the Kagom\'e plane. The best fit to the DM model is depicted
by the lines in Fig.~3(a) and describes the data very well.  We
reproduce not only the gaps at the zone center, but also the small
dispersion of the flat-mode.  This small dispersion is a result of a
weak $nnn$ interaction $J_2$. We note that $J_2$ is positive
(antiferromagnetic), which favors the experimentally observed ground
state.  The $D_z$ component of the DM vector also reinforces
selection of this ground state.  The ``zero energy mode'' is lifted
by an energy equal to the out-of-plane spin wave gap, consistent
with the spin rotations depicted in Fig.~1(a).  The gaps at the
$\Gamma$ point obtained numerically are in good agreement with the
analytic results given in Table~\ref{OMEGA}.  The DM model yields a
spin-canting angle of about 2.1$^{\rm o}$, in reasonable agreement
with the experimentally estimated value of 1.5(5)$^{\rm
o}$~\cite{estimate}.

The lines in Fig.~3(b) show the best fit to the CF model, which also
is in reasonable agreement with the data.  This model predicts a
canting angle of 0.8$^{\rm o}$, which is slightly smaller than, but
consistent with, the deduced experimental value.  However, the
least-squares goodness-of-fit to the spin wave spectrum is not as
good as that using the DM model.  The difference is most apparent in
the numerical results for the mode splitting at the zone center. The
CF model yields a relatively large splitting of about 0.71 meV for
the 7 meV mode at the $\Gamma$ point, whereas the data indicate that
this splitting is smaller than 0.4 meV.  The red line in Fig.~2(c)
shows the CF prediction for the energy scan at the zone center, and
we see that this does not describe the lineshape very well.  The DM
model, depicted by the blue line, describes the data better.
Moreover, as pointed out in Ref.~\cite{elhajal}, the single ion
anisotropy of the Fe$^{3+}$ ion is expected to be small since it
appears at second order in the spin-orbit coupling, whereas the DM
term appears at first order.

\begin{table}
\caption{\label{OMEGA}Spin wave energies at the zone center for the
DM model and the CF model. Here, $\tilde \omega \equiv \omega/S$, $J
\equiv J_1+J_2$, and $\theta_o \approx 20^{\rm o}$ is the oxygen
octahedra tilting angle.}\vspace{1mm}
\begin{tabular}{| c | c | c |}
\hline \hline
& DM Model & CF Model \\
\hline
$\tilde \omega_0$ & $\sqrt {12} |D_y|$ & $\sqrt{12J(E-D\sin^2(\theta_o) +E \cos^2(\theta_o))}$ \\
\hline
$\tilde \omega_{\pm} $ & $[3D_y^2 -6 \sqrt 3 JD_z]^{1/2}$ & $[6J(D+E)\cos(2\theta_o)]^{1/2}$ \\
& $\pm 5D_yD_z/J$ & $\pm (D+E) \sin(2\theta_o)/2$ \\
\hline \hline
\end{tabular}\vspace{-4mm}
\end{table}

Therefore, we believe the observed spin wave spectrum is most
naturally explained by a simple model which has only nearest and
next-nearest isotropic interactions plus the DM interaction.  The
obtained fit parameters (in meV) are $J_1=3.23(4)$, $J_2=0.11(1)$,
$|D_y|=0.218(5)$, and $D_z=-0.195(5)$. From a previous
study\cite{grohol_chiral}, a value for $J_1$ of 3.9(2)~meV was
obtained from a fit of the susceptibility to a high-temperature
series expansion result\cite{harris}.  The values of $J_1$ are in
reasonable agreement, and the agreement would be even closer if the
effects of $J_2$ and the DM term were taken into account in the
susceptibility fit.  As a further comparison, the susceptibility in
Ref.~\cite{grohol_chiral} indicated a value of $\Delta g/g \sim
0.06$, where $g$ is the free electron Land\'e factor and $\Delta g$
is its shift in the crystalline environment.  The magnitude of the
DM vector can be estimated from Moriya's calculation as
$|\vec{D}|/J_1 \sim \Delta g/g$~\cite{moriya}. From the current
study, we have $D_y/J_1 \sim |D_z|/J_1 \sim 0.06 \sim \Delta g/g$,
showing very good agreement between measurements of the spin
dynamics and the bulk thermodynamics.

Finally, from the analytic expressions for the spin gaps given in
Table~\ref{OMEGA}, we note that the in-plane gap is proportional to
$|D_y|$ while the out-of-plane gaps are proportional to $\sqrt{J_1
D_z}$. Since $J_1$ is large compared to other interactions, the
out-of-plane gap is significantly higher than the in-plane gap,
despite the similar magnitude of $D_y$ and $D_z$. This also suggests
that at high temperatures (even above $T_N$), the spins would feel
an easy-plane anisotropy and therefore display $XY$-like spin
dynamics at high temperatures.  This picture is consistent with the
previous neutron scattering measurements of the critical
fluctuations above $T_N$ which indeed have $XY$
symmetry\cite{grohol_chiral}.  That previous study also showed that
vector chiral order is apparent above $T_N$, consistent with the
presence of the DM term and the positive sign of $J_2$. Another
interesting aspect of the analytic results is that the spin wave
spectrum does not depend on the sign of $D_y$, which determines the
direction of the spin-canting relative to the in-plane order.  This
information on the canting cannot be determined from the currently
available data and would be important for testing microscopic
calculations of the DM interactions.

In summary, the spin wave spectrum of a Kagom\'e lattice
antiferromagnet has been measured using inelastic neutron
scattering. We observe a flat, lifted ``zero energy mode'' at
$\sim$7 meV, whose presence reflects the huge ground-state
degeneracy of the ideal Kagom\'e Heisenberg antiferromagnet. We have
also determined all of the relevant spin Hamiltonian parameters by
fitting our data to a Heisenberg model with the antisymmetric DM
interaction. This realization of the Kagom\'e antiferromagnet is
perhaps the best characterized geometrically frustrated spin system,
and, as such, would be useful for precise tests of theoretical
predictions. These results also highlight the importance of single
crystal measurements to accurately interpret the data taken on
powder samples of frustrated magnets.

We thank M.D. Lumsden, J.W. Lynn, and J.R.D. Copley for useful
discussions. The work at MIT was supported by the National Science
Foundation under award number DMR 0239377, and, in part, by the
MRSEC program under award number DMR 02-13282.  This work utilized
facilities supported in part by the NSF under Agreement No.
DMR-0454672.  ORNL is managed for the US DOE by UT-Battelle, LLC.
under Contract No. DE-AC05-00OR22725. \vspace{-6mm}

\bibliography{spin-waves}

\begin{thebibliography}{24}
\bibitem[{est()}]{estimate}
\bibinfo{note}{The spin canting angle was determined to be $0.65(6)^\circ$ at
  $T=50$~K in Ref.~[19].~~The low temperature canting angle is estimated to be
  $1.5(5)^\circ$ by extrapolating the ferromagnetic moment per Fe ion,
  $M_F(T)$, to $T=10$~K.}
\end{thebibliography}

\end{document}